\documentclass[aps,prl,reprint,superscriptaddress]{revtex4-2}

\usepackage{amsmath, amssymb, bm, xcolor}
\usepackage{graphicx}
\usepackage{comment}
\graphicspath{{figures/}}
\begin{document}

\title{Witnessing the Quantum Mpemba Effect with a Single  Observable}

\author{J. M. Z. Choquehuanca}
\affiliation{Instituto de F\'isica, Universidade Federal Fluminense, Av. Gal. Milton Tavares de Souza s/n, Gragoat\'a, 24210-346, Niter\'oi, RJ, Brazil}

\author{P. A. C. Obando}
\affiliation{Departamento de F\'isica and Centre for Bioinformatics and Photonics—CIBioFi, Universidad del Valle, 760032, Cali, Colombia}

\author{M. S. Sarandy}
\affiliation{Instituto de F\'isica, Universidade Federal Fluminense, Av. Gal. Milton Tavares de Souza s/n, Gragoat\'a, 24210-346, Niter\'oi, RJ, Brazil}

\author{F. M. de Paula}
\affiliation{Centro de Ci\^ encias Naturais e Humanas, Universidade Federal do ABC, Avenida dos Estados 5001, 09210-580, Santo Andr\'e, S\~ao Paulo, Brazil}
\date{\today}

\begin{abstract}

The Mpemba effect, in which a system farther from equilibrium relaxes faster than one closer to it, defies the intuition of nonequilibrium statistical mechanics. Here, we introduce a general framework to unambiguously witness this phenomenon based on a single observable defined for Markovian open quantum systems. The Mpemba observable bypasses full state tomography, offering utility for state preparation and thermal task optimization. We illustrate our method using local magnetization as a Mpemba signature, applying it to a qubit under generalized amplitude damping.

\end{abstract}

\maketitle

{\it Introduction.---} 
The Mpemba effect is a counterintuitive phenomenon known since ancient times, where a system at higher temperature relaxes to equilibrium faster than an identical system at lower temperature (for an extensive review, see Ref.~\cite{Teza2026}). 
Some decades ago, the Mpemba effect was rediscovered in the context of freezing water~\cite{Mpemba1969} and, more recently, it has been reframed within the landscape of relaxation dynamics in nonequilibrium statistical mechanics~\cite{Lu2017,Lasanta2017}.  
This foundational framework has been successfully generalized to the quantum domain, giving rise to the Quantum Mpemba Effect (QME)~\cite{Carollo2021, Kochsiek2022, Chatterjee2023}. In the QME, an open quantum system is swiftly driven to a final steady state following a time evolution dictated by the spectral properties of the Liouvillian superoperator. More specifically, initial state engineering can remove projections onto the slowest decaying modes of the dynamical superoperator, thereby enabling an exponentially accelerated path toward the steady state. In a perspective of quantum thermodynamics, this speed-up allows an initial ``hot" state, which is more distant from a final equilibrium state, to achieve equilibrium faster than an initial ``warm" state, which is closer to equilibrium at the beginning of evolution~\cite{Moroder2024}. 
Experimental investigations of the QME have recently been implemented through several architectures, such as trapped-ion experiments~\cite{Joshi2024,Shapira2024,Zhang2025} and nuclear magnetic resonance  platforms~\cite{Chatterjee2025,Schnepper2025}. These realizations show that anomalous relaxation is indeed an accessible feature of quantum thermalization. We also observe that even though the QME was originally proposed for Markovian open-system dynamics, it has also been extended to the non-Markovian regime, highlighting its potential for optimizing quantum thermal tasks~\cite{Strachan2025}.

Despite these advances, a significant challenge remains in the characterization of the QME by different metrics. The current literature presents a variety of possible characterizers (see, e.g., Ref.~\cite{Teza2026}), including relative entropy, trace distance, ergotropy, and nonequilibrium temperature. Crucially, these various measures often do not agree on either the existence or the crossing time for the QME to occur, leading to a variety of interpretations based on the chosen monotone. This lack of universality has prompted efforts toward  stricter definitions, such as those based on thermomajorization~\cite{Vu2025}. Experimentally, the identification of the Mpemba crossover time is also notoriously difficult because standard distance-based characterizers typically require full quantum state tomography. This is a resource-intensive process, which demands the reconstruction of the entire density matrix through exhaustive measurement sets. Indeed, the dependence on quantum tomography makes real-time tracking of the QME crossing point nearly impossible in a real-world environment, where only limited observables may be accessible.

To address these limitations, we propose a directly measurable observable, the {\it Mpemba observable}, which serves as a rigorous ``witness'' for the QME. By constructing a Hermitian operator whose expectation value is mathematically guaranteed to exhibit the exact characteristic crossing time of an arbitrarily chosen Mpemba measure (e.g., the trace distance), we provide a method to identify the effect without the need for full state reconstruction. This approach provides a practical tool for detecting the dynamical signature of the Mpemba effect, potentially facilitating its use as a resource for state preparation and thermal task optimization. We discuss how to construct the Mpemba observable in a general Markovian open-system evolution setup. Moreover, we illustrate its behavior by adopting distinct characterizers. Specifically, we consider a qubit evolving under a Markovian generalized amplitude damping (GAD) quantum channel, analyzing the QME by both a geometric approach, adopting the trace distance as a QME measure, and a thermal approach, using the ergotropic  temperature (recently introduced in Ref.~\cite{Choquehuanca2025}) to identify the QME. In particular, we analytically obtain the precise direction of the qubit local magnetization based on a single parameter, providing the time-crossing signature of the QME.

\textit{Markovian open quantum dynamics.}---
We begin by setting up the essential features 
of open quantum systems evolving under Markovian dynamics~\cite{Lindblad1976,
Gorini1976}. Let  $\rho(t)$ denote a density operator acting on a $d$-dimensional Hilbert space ${\cal H}$ and evolving according to a convolutionless master equation,
\begin{equation}
  \frac{d\rho}{dt} = \mathcal{L}[\rho],
  \label{eq:lindblad}
\end{equation}
where the Liouvillian $\mathcal{L}$ (hereafter  the Lindblad superoperator) is time-independent and generates a completely positive trace-preserving (CPTP) dynamics~\cite{Breuer2002}. In addition, we assume $\rho(t)$ exhibits a unique steady state, 
$\rho_{\mathrm{eq}} = \lim_{t \to \infty} \rho(t)$, 
with $\mathcal{L}[\rho_{\mathrm{eq}}] = 0$.

The formal solution to the quantum master equation is given by
$\rho(t) = e^{t\mathcal{L}}[\rho_0]$, where $\rho_0$ is the density operator at time $t=0$. Assuming the generator $\mathcal{L}$ 
is diagonalizable (see, e.g. Refs.~\cite{Sarandy2005,Santos2020} for more general cases), the spectral
decomposition of $\mathcal{L}$ and its adjoint $\mathcal{L}^\dagger$ is
determined by the right $R_i$ and left $L_i$  eigenmatrices,
\begin{equation}
  \mathcal{L}[R_i] = \lambda_i R_i,
  \qquad
  \mathcal{L}^\dagger[L_i] = \lambda_i^* L_i,
  \label{eq:eigeqs}
\end{equation}
where the complex numbers $\lambda_i$ are the eigenvalues of the Lindblad superoperator. The matrices $R_i$ and $L_i$ in Eq.~\eqref{eq:eigeqs}, satisfy the biorthonormality condition $\mathrm{Tr}\!\left(L_i^\dagger R_j\right) = \delta_{ij}$.
Since $\mathcal{L}$ implements a CPTP map, all of its eigenvalues have a non-positive real part, $\mathrm{Re}(\lambda_k) \leq 0$. Furthermore, trace preservation enforces that at least one eigenvalue vanishes; we take $\lambda_0 = 0$. Assuming $\lambda_0$ is non-degenerate, the steady state is unique, given by the right eigenmatrix $R_0$~\cite{Carollo2021}. The remaining eigenvalues in Eq.~(\ref{eq:eigeqs}) thus take the form $\lambda_i = -\gamma_i + i\omega_i$, with $\gamma_i \geq 0$.
Now, consider $\mathcal{L}=\mathcal{L}_u+\mathcal{L}_d$, with $\mathcal{L}_u$ and $\mathcal{L}_d$ describing the unitary and dissipative contributions of $\mathcal{L}$, respectively. Assuming that $\mathcal{L}$ implements a Davies map~\cite{Davies1974}, we have $\left[\mathcal{L}_u,\mathcal{L}_d\right]=0$. Working in the {\it{interaction picture}}, we write $\rho(t)=e^{-\mathcal{L}_u t}\,e^{(\mathcal{L}_u+\mathcal{L}_d) t}[\rho_0]=e^{\mathcal{L}_d t}[\rho_0]$, which removes the oscillatory contributions (the imaginary parts of the eigenvalues) present in the Schr\"odinger-picture solution.
The general solution of Eq.~\eqref{eq:lindblad} then reads $\rho(t) = \sum_{i=0}^{d^2-1} \alpha_i\, e^{-\gamma_i t} R_i$,
where the coefficients $\alpha_i= \mathrm{Tr}\!\left(L_i^\dagger \rho_0\right)$ encode the overlap of the initial state
$\rho_0$ with the $i$-th decay mode. 
 In this scenario, we obtain
\begin{equation}
  \rho(t)  = \rho_{\mathrm{eq}} + \sum^{d^2-1}_{i = 1} \alpha_i\, e^{-\gamma_i t} R_i,
  \label{eq:sol_real}
\end{equation}
where we have used $\alpha_0 = 1$, $\gamma_0 = 0$, $R_0 = \rho_{\mathrm{eq}}$, and $L_0 = \mathbb{I}$,
with $\mathbb{I}$ denoting the identity operator on ${\cal H}$.\\


\textit{Mpemba observable.}---The QME generally manifests as
a crossing at a finite time $t_M$ between the values of a
quantity for two systems evolving toward the same equilibrium state.
This \emph{Mpemba time} encodes the competition between the decay modes of the Lindblad superoperator and can be used to determine the structure of a corresponding observable. We now formalize this picture by defining the Mpemba observable 
$\mathcal{M}$ as a Hermitian operator whose expectation value witnesses this
characteristic time. Mathematically, given two identical systems $A$ and $B$ prepared in
initial states $\rho_{A(B)}$ and thermalizing under the same
dynamics, a well-defined Mpemba observable must satisfy two conditions:
\begin{enumerate}
    \item {Hermiticity:} $\mathcal{M} = \mathcal{M}^\dagger$,
    \item {Crossing condition:} $  M_A(t_M) = M_B(t_M)$,
\end{enumerate}
where $M_{A(B)}(t) = \mathrm{Tr}(\mathcal{M}\,\rho_{A(B)}(t))$ denotes the
expectation value of $\mathcal{M}$ in the time-evolved state of each system. By using Eq.~\eqref{eq:sol_real}, the expectation value
of $\mathcal{M}$ at time $t$ takes the form
\begin{equation}
  M(t) = \mathrm{Tr}\!\left(\mathcal{M}\,\rho(t)\right)
       = M_{\mathrm{eq}} + \sum_{i=1}^{d^2-1} \alpha_i\, e^{-\gamma_i t}\, m_i,
  \label{eq:expectation}
\end{equation}
where we have introduced the mode-resolved elements
$m_i \equiv \mathrm{Tr}(\mathcal{M}\, R_i)$ ($i=0,\cdots,d^2-1$) and the equilibrium expectation
value $M_{\mathrm{eq}} \equiv \mathrm{Tr}(\mathcal{M}\,\rho_{\mathrm{eq}})=m_0$. As a consequence of the biorthonormality relation, $\mathcal{M}$
admits the expansion $\mathcal{M} = \sum_{i=0}^{d^2-1} m_i\, L_i^\dagger$
on the basis of the left eigenmatrices of $\mathcal{L}$. Condition 1 imposes the constraint
\begin{equation}
  \sum_{i=0}^{d^2-1} \left( m_i\, L_i^\dagger - m_i^*\, L_i \right) = 0
  \label{eq:herm_condition}
\end{equation}
and further ensures the spectral decomposition $\mathcal{M} = \sum_{i=1}^{d} \mu_i\, P_i$, 
where $\{\mu_i\}$ are the eigenvalues of $\mathcal{M}$,
$P_i = |\mu_i\rangle\langle\mu_i|$ the corresponding projectors, and
$\{|\mu_i\rangle\}$ defines the \emph{Mpemba basis} associated with
$\mathcal{M}$. Thus, Eq.~\eqref{eq:herm_condition} establishes a direct link
between the measurement basis of $\mathcal{M}$ and the dynamical structure of
the Lindblad superoperator: the coefficients $\{m_i\}$ simultaneously encode the spectral
content of the observable and its overlap with the decay modes that govern
the relaxation. Condition 2, in turn, requires that the difference in expectation values $M_A(t) - M_B(t)$ vanish at
$t = t_M$. From Eq.~\eqref{eq:expectation}, this translates into
\begin{equation}
  \sum_{i=1}^{d^2-1} \Delta\alpha_i\, m_i\, e^{-\gamma_i t_M} =0
  \label{eq:mpemba_cond}
\end{equation}
where $\Delta\alpha_i \equiv \alpha_{Ai} - \alpha_{Bi} $ quantifies the differential overlap of the two initial
states with the $i$-th decay mode of the Lindblad superoperator. The observable
$\mathcal{M}$ must therefore be chosen such that its mode projections
$\{m_i\}$ produce a weighted cancelation of these overlaps in a finite time $t_M$.

When several eigenmodes share the same decay rate $\gamma_i$, it is convenient to
group them into a single effective contribution. Denoting by $g_i$ the
degeneracy of the $i$-th decay rate, we require  $m_i^{(j)}\equiv \mathrm{Tr}(\mathcal{M}R_i^{(j)})=\mathrm{Tr}(\mathcal{M}R_i^{(1)})=m_i$ $(\forall j)$ and define the aggregated overlap $\tilde{\alpha}_i = \sum_{j=1}^{g_i} \alpha_i^{(j)}$,
where $\alpha_i^{(j)} = \mathrm{Tr}(L_i^{(j)\dagger}\rho_0)$ is the overlap
of the initial state with the $j$-th eigenmode in the degenerate subspace.
Eq.~\eqref{eq:mpemba_cond} then takes the form
\begin{equation}
  \sum_{i=1}^{d^2-1} \Delta\tilde{\alpha}_i\, m_i\, e^{-\gamma_i t_M} = 0,
  \label{eq:deg_condition}
\end{equation}
which reduces to Eq.~\eqref{eq:mpemba_cond} in the non-degenerate case
$g_i = 1$ for all $i$. Furthermore, in many physically relevant situations, the relaxation is dominated at
intermediate and long times by two modes with well-separated decay rates
$\gamma_s < \gamma_f$, where the subscripts $s$ and $f$ label the slower and
faster decaying contributions, respectively. Retaining only these two modes,
the crossing condition~\eqref{eq:deg_condition} reduces to
\begin{equation}
  \Delta\tilde{\alpha}_s\, m_s\, e^{-\gamma_s t_M}
  +
  \Delta\tilde{\alpha}_f\, m_f\, e^{-\gamma_f t_M}
  = 0,
  \label{eq:two_mode}
\end{equation}
where $\Delta\tilde{\alpha}_i = \tilde{\alpha}_{Ai} - \tilde{\alpha}_{Bi}$ is
the differential aggregated overlap between the two initial states and the
$i$-th mode group. By introducing the ratio
\begin{equation}
  \eta \equiv -\frac{\Delta\tilde{\alpha}_s}{\Delta\tilde{\alpha}_f}\,
         e^{(\gamma_f - \gamma_s)\,t_M},
  \label{eq:eta}
\end{equation}
Eq.~\eqref{eq:two_mode} yields a simple constraint relating the projections of the Mpemba observable onto the two dominant
modes:
\begin{equation}
  m_f = \eta\, m_s.
  \label{eq:mr_ml}
\end{equation}
This relation constitutes the central design equation for the construction of $\mathcal{M}$ in the two-mode regime. It encodes the Mpemba crossing in a single dimensionless dynamical scalar $\eta$ (assumed to be finite and nonvanishing), entirely determined by the initial conditions.
\medskip

\textit{Qubit under Generalized Amplitude Damping}.---  Let us consider a qubit under a GAD process, whose dissipative part is given by  $\mathcal{L}_d[{\bullet}] = \gamma_- D_-[{\bullet}] + \gamma_+ D_+[{\bullet}]$~\cite{Sapui2026}.
The Hamiltonian is given by $H=-\vec{\omega}\cdot\vec{\sigma}$, with $\vec{\omega}=(0,0,\omega)$ denoting an effective magnetic field along the $z$-axis (by convection) and $\vec{\sigma}=(\sigma_x,\sigma_y,\sigma_z)$ the Pauli operator vector. The dissipative term depends on $D_\mp[{\bullet}] \equiv \sigma_\mp \,(\bullet)\, \sigma_\pm- \tfrac{1}{2}\left\{\sigma_\pm \sigma_\mp,\, {\bullet}\right\}$, where $\sigma_{\pm}=(\sigma_x\pm i\sigma_y)/2$ are the ladder operators, and involves the absorption and emission decay rates $\gamma_\pm$, whose ratio is governed by the Boltzmann factor $\gamma_+/\gamma_- = e^{-2\omega/k_B T_{\mathrm{eq}}}$, where $2\omega$ is the energy gap, $k_B$ the Boltzmann constant, and $T_{\mathrm{eq}}$ the equilibrium temperature. In this case, the qubit dynamics is described by the density operator in Eq.\eqref{eq:sol_real} with 
\begin{align*}
    \{R_i\} & = \{(\mathbb{I}+z_{\mathrm{eq}}\sigma_z)/2,\,\sigma_ x/2,\, \sigma_y/2,\, \sigma_z/2\} ,\\
    \{L_i\} & = \{\mathbb{I},\,\sigma_ x,\, \sigma_y,\, \sigma_z-z_{\mathrm{eq}}\mathbb{I}\}, \\
    \{\gamma_i\} & = \{0,\,\gamma/2,\, \gamma/2,\, \gamma\}, \\
    \{\alpha_i\} & = \{1,\,x_0,\, y_0,\, z_0-z_{\mathrm{eq}}\},
\end{align*}
where $\vec{r}_{0}=(x_{0},y_{0},z_{0})$ and $\vec{r}_{\mathrm{eq}}=(0,0,z_{\mathrm{eq}})$, with  $z_{\mathrm{eq}} =\tanh\left(2\omega/{k_BT_{\mathrm{eq}}}\right)$, represent the Bloch vectors associated with $\rho_0$ and $\rho_{\mathrm{eq}}$, respectively, and $\gamma \equiv \gamma_+ + \gamma_-$ defines the total rate.  Equivalently, the solution can be written in the form $\rho(t)=[\mathbb{I}+\vec{r}(t)\cdot\vec{\sigma}]/2$ with the Bloch vector evolving as $\vec{r}(t) =(x_{0}\,e^{-\gamma t/2},\,y_{0}\,e^{-\gamma t/2},\, z_{\mathrm{eq}}+z_{0}^{\mathrm{eq}}\,e^{-\gamma t})$,
where $z_{0}^{\mathrm{eq}} \equiv z_{0}-z_\mathrm{eq}$. The corresponding Mpemba observable expands in the left-eigenvector basis as $\mathcal{M} 
  = m_0 I + m_1\,\sigma_x + m_2\,\sigma_y + m_3\left(\sigma_z - z_{\mathrm{eq}}I\right)$.
Since $L_i = L_i^\dagger$, condition~\eqref{eq:herm_condition} is satisfied for $m_i \in \mathbb{R}$. Furthermore, there are effectively two modes: a degenerate coherence mode (the slow mode with rate $\gamma_s=\gamma/2$) and an incoherent mode (the fast mode with rate $\gamma_f=\gamma$). Consequently, condition ~\eqref{eq:mr_ml} imposes $m_3 = \eta\, m_2$, with $m_2 = m_1$, and 
\begin{equation}
  \eta=\eta _0e^{\gamma t_M/2},\,\,\,\,\text{where}\,\,\,\,\,\eta_0\equiv \frac{x_{A0}-x_{B0}+y_{A0}-y_{B0}}{z_{B0}-z_{A0}},
  \label{eq:eta_qubit}
\end{equation}
with $\eta_0$ nonvanishing and finite. By adopting $\mathrm{Tr}\, (\mathcal M)=0$, we take $m_0 = z_{\mathrm{eq}}m_3$. In addition, we require a normalized observable $\mathcal M$, {\it{i.e.}} $\mathrm{Tr}\, (\mathcal M^2)=1$, which is achieved by setting 
$m_1=m_2 = (2 + \eta^2)^{-1/2}$. Hence, the Mpemba observable reads
\begin{equation}
  \mathcal{M} = \hat{m}\cdot\vec{\sigma},
  \qquad
  \hat{m} = \frac{1}{\sqrt{2+\eta^2}}\,(1,\,1,\,\eta),
  \label{eq:mpemba_obs_final}
\end{equation}
where $\hat{m}$ is the \textit{Mpemba vector}, a unit vector which defines the measurement
direction (basis) on the Bloch sphere. The spectral decomposition of the Mpemba observable is remarkably simple: $\mathcal{M} = \mu_+ P_+ + \mu_- P_-$, where $\mu_\pm = \pm 1 $ denote the eigenvalues and $P_\pm = (\mathbb{I} \pm \hat{m}\cdot\vec{\sigma})/{2}$ the corresponding projectors. Furthermore, the \textit{Mpemba mean} (i.e., the expectation value of $\mathcal{M}$) is 
\begin{equation}
  M(\rho) 
  = p_+ - p_-= \hat{m}\cdot\vec{r},
  \label{eq:expectation_val}
\end{equation}
where $ p_\pm = \mathrm{Tr}(P_\pm\,\mathcal{M}) = (1 \pm \hat{m}\cdot\vec{r})/{2}$ are the probabilities associated with the outcomes $\pm 1$. Therefore, by preparing $N \gg 1$ identical copies of $\rho$ and performing projective measurements along the direction $\hat{m}$ in the rotating frame, counting the number $N_{\pm}$ of outcomes $\pm 1$, one obtains experimentally the probabilities $ p_\pm=N_{\pm}/N $ and,
consequently, the Mpemba mean.
\medskip

\begin{figure}[!h]
  \centering
  \includegraphics[width=\columnwidth]{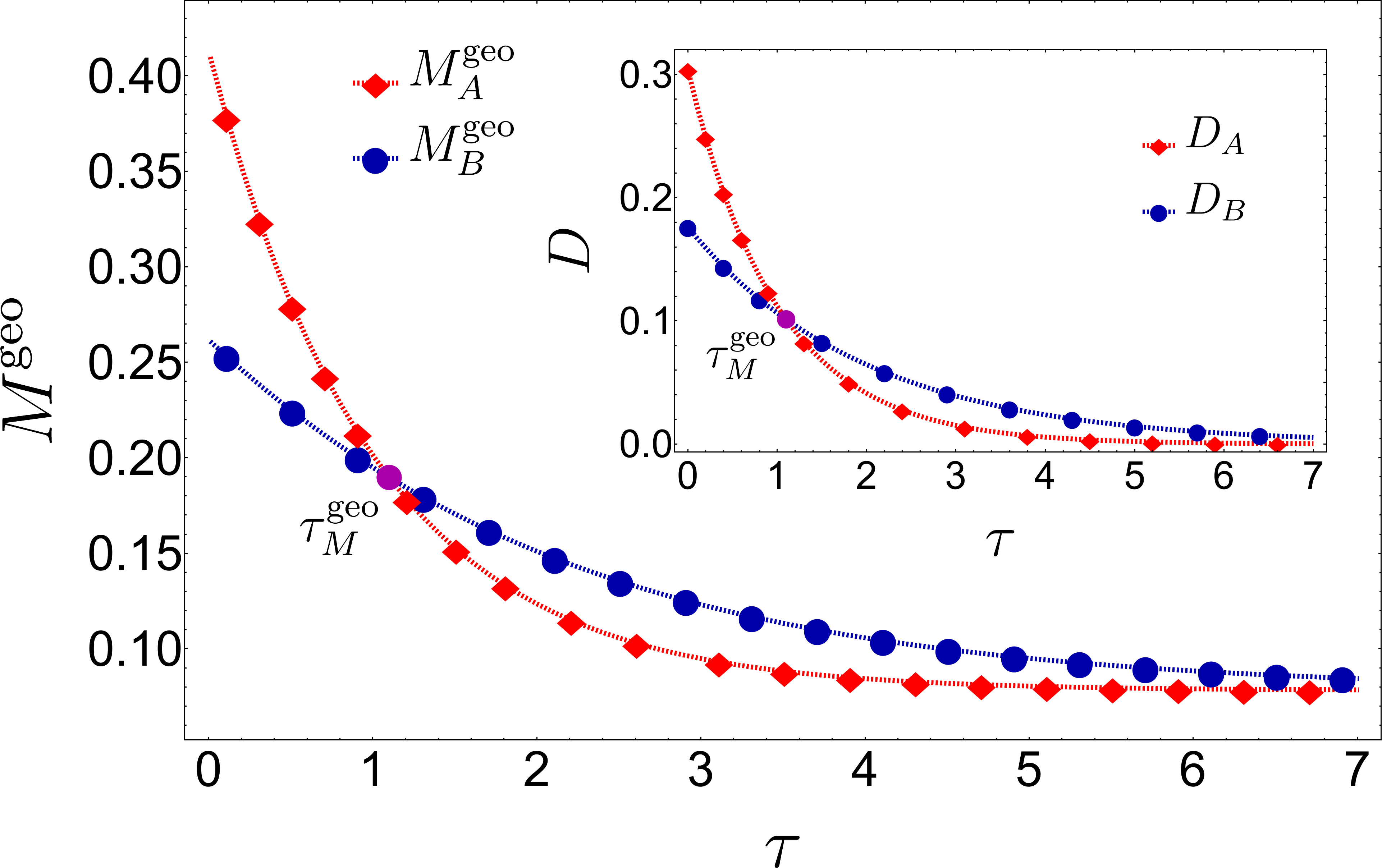}
  \caption{(Color online) Geometric Mpemba means $M_{A}^{\text{geo}}$ and $M_{B}^{\text{geo}}$ as functions of the dimensionless time $\tau=\gamma t$ for $\vec{r}_{A0}=(0,0,0.75)$, $\vec{r}_{B0}=(0.35,0,0.10)$, and $\vec{r}_{\mathrm{eq}}=(0,0,0.14)$. Inset: Trace distances $D_A$ and $D_B$ as functions of $\tau$ for the same initial conditions. The crossing occurs at geometric Mpemba time $\tau_{M}^{\mathrm{geo}}=1.1$.}
  \label{fig:distances}
\end{figure}

\textit{Geometric Quantum Mpemba Effect}.--- To illustrate our method, we first analyze the QME from a geometric perspective, adopting the \textit{trace distance} $D(t)$ to quantify the degree of separation between a qubit and its equilibrium state under a Markovian GAD process. The trace distance between the states $\vec{r}(t)$ and $\vec{r}_{eq}$ is given by \cite{nielsen2000}
\begin{equation}
  D(t) = \frac{1}{2}\bigl|\vec{r}(t) - \vec{r}_\mathrm{eq}\bigr|.
  \label{eq:trace_distance}
\end{equation}
The \textit{geometric Mpemba time} $t_{M}^{\mathrm{geo}}$ is defined by the condition
that two initially distinct states $\vec{r}_A$ and $\vec{r}_B$ reach equal trace distances from
equilibrium simultaneously: $D_{A}\left(t_{M}^{\mathrm{geo}}\right) = D_{B}\left(t_{M}^{\mathrm{geo}}\right)$. Solving for the GAD process yields a unique solution
\begin{equation}
  t_{M}^{\mathrm{geo}} =\frac{1}{\gamma} \ln\left(
\frac{z_{A0}^{\mathrm{eq}\,2} - z_{B0}^{\mathrm{eq}\,2}}
{c_{B0}^{2} - c_{A0}^{2}}
\right),
  \label{eq:mpemba_time}
\end{equation}
where $c_{k0}=\sqrt{x_{k0}^2+y_{k0}^2}$ provides the initial \textit{$l_1$-norm of coherence} \cite{Baumgratz2014} of the qubit $k$ ($k=A$ or $B$). A crossing at a positive time requires that the argument of the logarithm is greater than one and finite. In particular, the qubits, $A$ and $B$, must differ in their initial energies and quantum
coherences. Quantum coherence acts as a \emph{brake} on relaxation: larger coherence slows down the approach to equilibrium, thereby enabling the geometric QME. Figure~\ref{fig:distances} shows the geometric Mpemba means $M_A^{\text{geo}}$ and  $M_B^{\text{geo}}$ as well as their respective trace distances $D_{A}$ and
$D_{B}$ (inset) as functions of time. The state $B$ starts closer to equilibrium, but is overtaken by the state $A$, which relaxes faster due to its incoherent nature. Crucially, the crossings in both panels occur at the same
instant of time $t_{M}^{\mathrm{geo}}$, with the geometric Mpemba mean inheriting the QME from the trace distance.

\medskip

\textit{Thermal Quantum Mpemba Effect}.---  As a second application, we discuss the QME via a thermal approach, employing a measure of non-equilibrium temperature associated with a recently introduced ergotropic formulation of quantum thermodynamics \cite{Choquehuanca2025}. Within this formulation, heat is linked to changes in the von Neumann entropy, whereas work depends on variations in both ergotropy and Hamiltonian. In analogy with classical equilibrium thermodynamics, the inverse of the \textit{ergotropic temperature} is defined as the rate of change of von Neumann entropy with respect to the internal energy at zero ergotropic work. For an arbitrary qubit state, $1/T=(\partial S/\partial U)_{\vec{\omega},\,\mathcal{E}}$,
 where $dS = -k_B \tanh^{-1}r\,dr$ denotes an entropy change, $U=-\vec{\omega}\cdot\vec{r}$ the internal energy, and $\mathcal{E}= \omega r + U$ the ergotropy. Assuming a time-independent Hamiltonian, the temperature of a qubit in a state $\vec{r}(t)$ can be written as
\begin{equation}
T(t) = \frac{\omega}{k_{B}\tanh^{-1}\,r(t)}.
  \label{eq:qubit_temperature}
\end{equation}
The {\it{thermal Mpemba time}} $t_{M}^{\mathrm{ther}}$ is defined by the condition
that two initially distinct states $\vec{r}_A$ and $\vec{r}_B$ reach equal ergotropic temperature simultaneously: $T_{A}\left(t_{M}^{\mathrm{ther}}\right) = T_{B}\left(t_{M}^{\mathrm{ther}}\right)$.
Solving for the GAD process yields
\begin{equation}
  t_{M}^{\mathrm{ther}} =\frac{1}{\gamma} \ln\left(
\frac{z_{A0}^{\mathrm{eq}\,2} - z_{B0}^{\mathrm{eq}\,2}}
{c_{B0}^{2} - c_{A0}^{2}-2z_{\text{eq}}(z_{A0}^{\text{eq}}-z_{B0}^{\text{eq}})}
\right).
  \label{eq:mpemba_time}
\end{equation}

\begin{figure}[!h]
  \centering
  \includegraphics[width=\columnwidth]{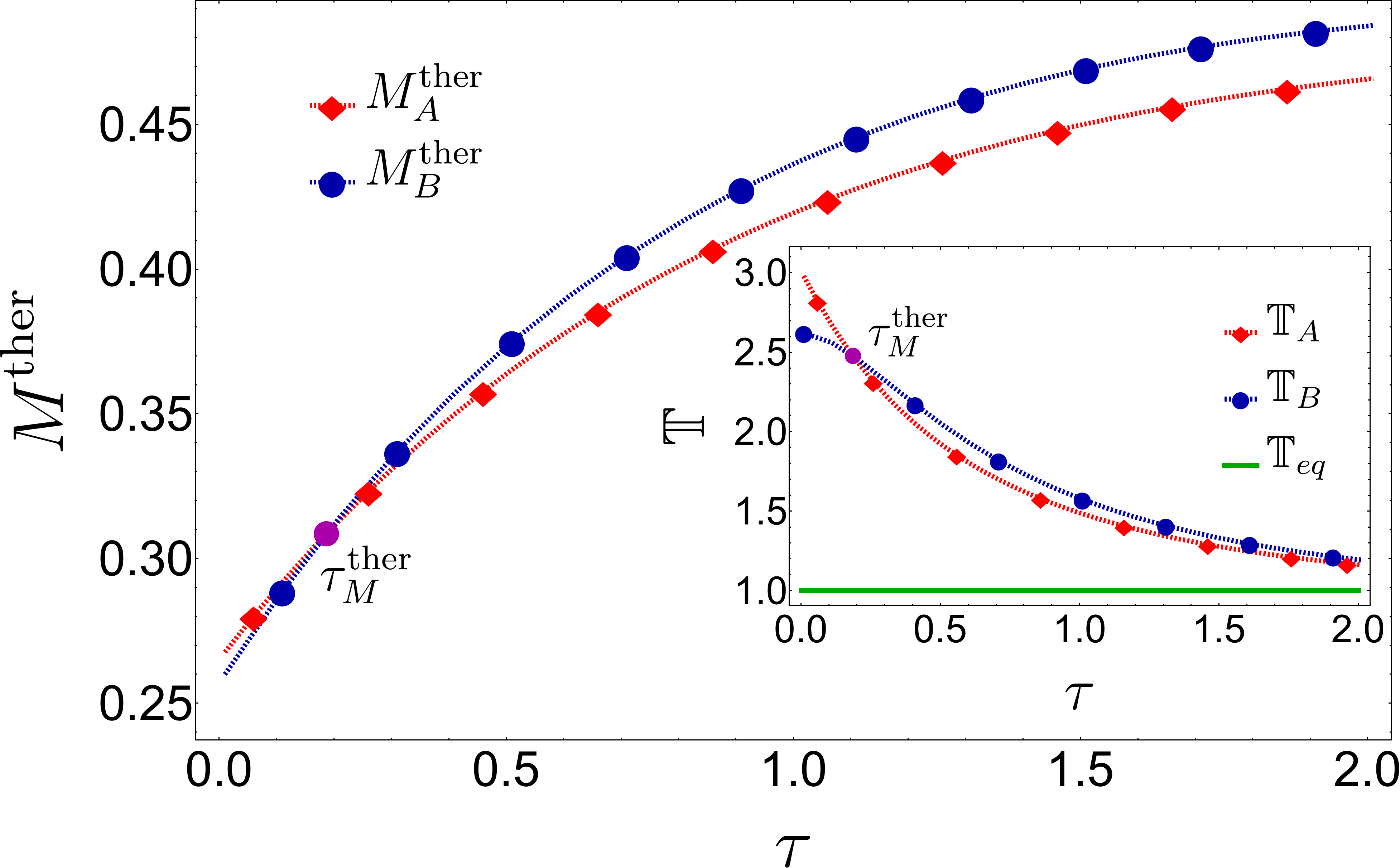}
  \caption{(Color online) Thermal Mpemba means $M_{A}^{\text{ther}}$ and $M_{B}^{\text{ther}}$ as functions of the dimensionless time $\tau=\gamma t$ for $\vec{r}_{A0}=(0.20,0,0.25)$, $\vec{r}_{B0}=(0.35,0,0.10)$, and $\vec{r}_{\mathrm{eq}}=(0,0,0.76)$. Inset: Dimensionless ergotropic temperatures $\mathbb{T}_A=k_BT_A/\omega$ and $\mathbb{T}_B=k_BT_B/\omega$ as functions of $\tau$ for the same initial conditions. The crossing occurs at thermal Mpemba time $\tau_{M}^{\mathrm{ther}}=0.19$.}
  \label{fig:2}
\end{figure}

\begin{figure}[!h]
  \centering
  \includegraphics[width=\columnwidth]{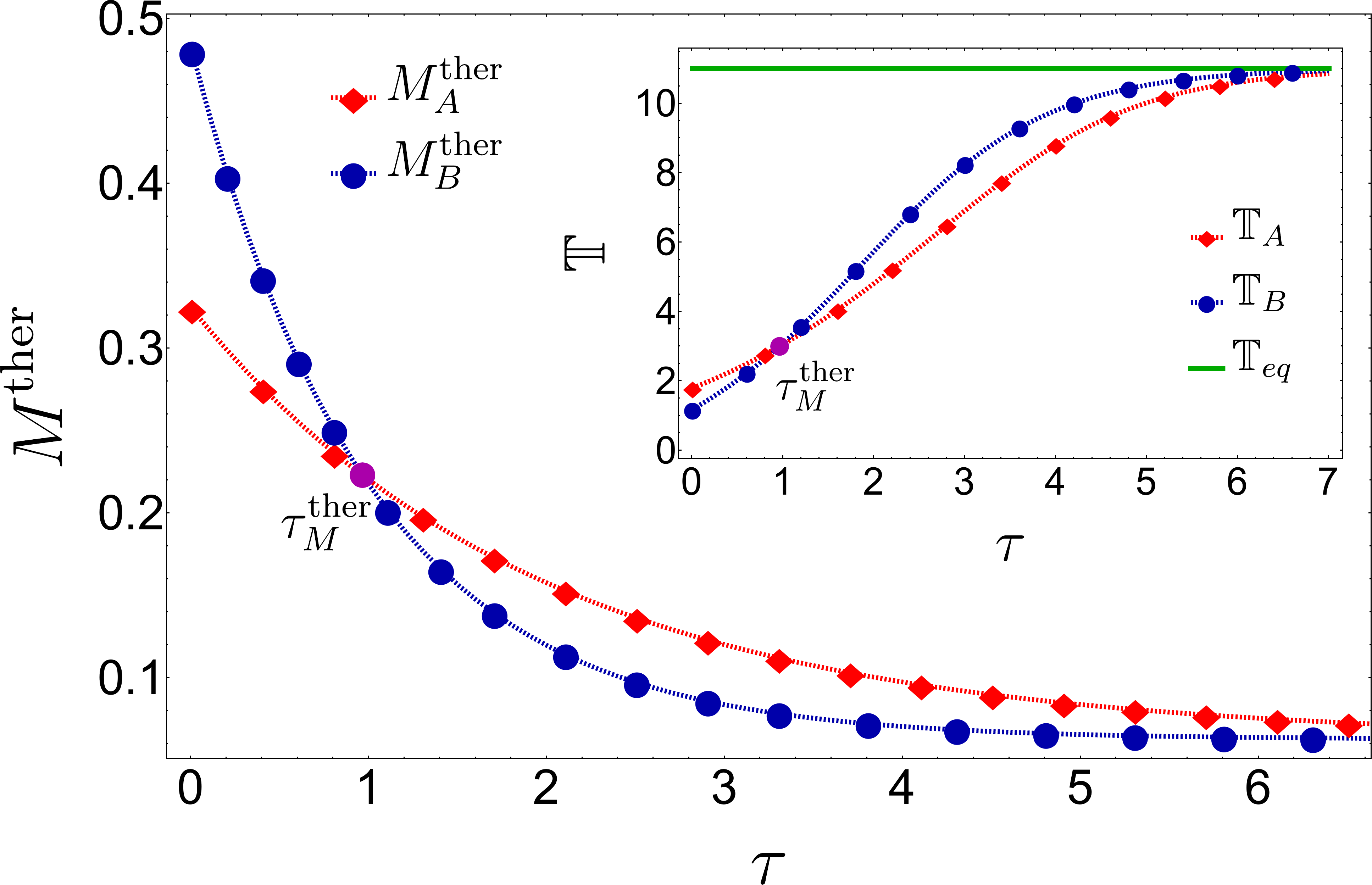}
  \caption{(Color online) Thermal Mpemba means $M_{A}^{\text{ther}}$ and $M_{B}^{\text{ther}}$ as functions of the dimensionless time $\tau=\gamma t$ for $\vec{r}_{A0}=(0.5,0,0.1)$, $\vec{r}_{B0}=(0,0,0.7)$, and $\vec{r}_{\mathrm{eq}}=(0,0,0.091)$. Inset: Dimensionless ergotropic temperatures $\mathbb{T}_A=k_BT_A/\omega$ and $\mathbb{T}_B=k_BT_B/\omega$ as functions of $\tau$ for the same initial conditions. The crossing occurs at thermal Mpemba time $\tau_{M}^{\mathrm{ther}}=0.97$.}
  \label{fig:3}
\end{figure}

Note that an additional term appears in the denominator of $t_M^{\text{ther}}$ compared to $t_M^{\text{geo}}$, which is negligible in the high-temperature limit ($z_{\text{eq}}\approx0$). Consequently, the emergence of \textit{thermal} QME does not necessarily depend on distinct initial quantum coherences. In case $c_{B0}=c_{A0}$, the requirement is a positive total energy (non-zero initial total incoherent ergotropy \cite{Francica2020,Choquehuanca2024}): $z_{A0}+z_{B0}<0$. Figure~\ref{fig:2} displays the thermal Mpemba means $M_A^{\text{ther}}$ and  $M_B^{\text{ther}}$ as well as their respective ergotropic temperatures $T_{A}$ and
$T_{B}$ as functions of time. The temperature of state $B$ starts closer to the equilibrium temperature, but is overtaken by the temperature of state $A$. As expected, the crossings in both panels coincide at $t_{M}^{\mathrm{ther}}$, demonstrating that the thermal Mpemba mean inherits the thermal QME of the ergotropic temperature. We also illustrate the inverse QME in Figure~\ref{fig:3}, where the temperatures increase with time.

{\it Conclusion.---}
We have introduced a general framework for characterizing the QME through a single observable. By exploiting the spectral properties of the dynamical generator, this approach significantly reduces complexity by avoiding quantum tomography. Crucially, the Mpemba observable is agnostic to the underlying measure used to achieve relaxation advantage. This has been illustrated by analyzing a qubit under GAD, demonstrating utility across both geometric (trace distance) and thermal (ergotropic temperature) characterizers. Natural extensions include scaling to many-body systems, non-Markovian effects, and experimental realizations.

{\it Acknowledgments.---} 
M.S.S. is supported by Conselho Nacional de Desenvolvimento Cient\'{\i}fico e Tecnol\'ogico (CNPq) (grant number 303836/2024-5).

\vspace{-0.6cm}


\begin{thebibliography}{27}%
\makeatletter
\providecommand \@ifxundefined [1]{%
 \@ifx{#1\undefined}
}%
\providecommand \@ifnum [1]{%
 \ifnum #1\expandafter \@firstoftwo
 \else \expandafter \@secondoftwo
 \fi
}%
\providecommand \@ifx [1]{%
 \ifx #1\expandafter \@firstoftwo
 \else \expandafter \@secondoftwo
 \fi
}%
\providecommand \natexlab [1]{#1}%
\providecommand \enquote  [1]{``#1''}%
\providecommand \bibnamefont  [1]{#1}%
\providecommand \bibfnamefont [1]{#1}%
\providecommand \citenamefont [1]{#1}%
\providecommand \href@noop [0]{\@secondoftwo}%
\providecommand \href [0]{\begingroup \@sanitize@url \@href}%
\providecommand \@href[1]{\@@startlink{#1}\@@href}%
\providecommand \@@href[1]{\endgroup#1\@@endlink}%
\providecommand \@sanitize@url [0]{\catcode `\\12\catcode `\$12\catcode
  `\&12\catcode `\#12\catcode `\^12\catcode `\_12\catcode `\%12\relax}%
\providecommand \@@startlink[1]{}%
\providecommand \@@endlink[0]{}%
\providecommand \url  [0]{\begingroup\@sanitize@url \@url }%
\providecommand \@url [1]{\endgroup\@href {#1}{\urlprefix }}%
\providecommand \urlprefix  [0]{URL }%
\providecommand \Eprint [0]{\href }%
\providecommand \doibase [0]{https://doi.org/}%
\providecommand \selectlanguage [0]{\@gobble}%
\providecommand \bibinfo  [0]{\@secondoftwo}%
\providecommand \bibfield  [0]{\@secondoftwo}%
\providecommand \translation [1]{[#1]}%
\providecommand \BibitemOpen [0]{}%
\providecommand \bibitemStop [0]{}%
\providecommand \bibitemNoStop [0]{.\EOS\space}%
\providecommand \EOS [0]{\spacefactor3000\relax}%
\providecommand \BibitemShut  [1]{\csname bibitem#1\endcsname}%
\let\auto@bib@innerbib\@empty
\bibitem [{\citenamefont {Teza}\ \emph {et~al.}(2026)\citenamefont {Teza},
  \citenamefont {Bechhoefer}, \citenamefont {Lasanta}, \citenamefont {Raz},\
  and\ \citenamefont {Vucelja}}]{Teza2026}%
  \BibitemOpen
  \bibfield  {author} {\bibinfo {author} {\bibfnamefont {G.}~\bibnamefont
  {Teza}}, \bibinfo {author} {\bibfnamefont {J.}~\bibnamefont {Bechhoefer}},
  \bibinfo {author} {\bibfnamefont {A.}~\bibnamefont {Lasanta}}, \bibinfo
  {author} {\bibfnamefont {O.}~\bibnamefont {Raz}},\ and\ \bibinfo {author}
  {\bibfnamefont {M.}~\bibnamefont {Vucelja}},\ }\bibfield  {title} {\bibinfo
  {title} {Speedups in nonequilibrium thermal relaxation: {Mpemba} and related
  effects},\ }\href {https://doi.org/10.1016/j.physrep.2025.10.009} {\bibfield
  {journal} {\bibinfo  {journal} {Phys. Reports}\ }\textbf {\bibinfo {volume}
  {1164}},\ \bibinfo {pages} {1} (\bibinfo {year} {2026})}\BibitemShut
  {NoStop}%
\bibitem [{\citenamefont {Mpemba}\ and\ \citenamefont
  {Osborne}(1969)}]{Mpemba1969}%
  \BibitemOpen
  \bibfield  {author} {\bibinfo {author} {\bibfnamefont {E.~B.}\ \bibnamefont
  {Mpemba}}\ and\ \bibinfo {author} {\bibfnamefont {D.~G.}\ \bibnamefont
  {Osborne}},\ }\bibfield  {title} {\bibinfo {title} {Cool?},\ }\href
  {https://doi.org/10.1088/0031-9120/4/3/312} {\bibfield  {journal} {\bibinfo
  {journal} {Phys. Educ.}\ }\textbf {\bibinfo {volume} {4}},\ \bibinfo {pages}
  {172} (\bibinfo {year} {1969})}\BibitemShut {NoStop}%
\bibitem [{\citenamefont {Lu}\ and\ \citenamefont {Raz}(2017)}]{Lu2017}%
  \BibitemOpen
  \bibfield  {author} {\bibinfo {author} {\bibfnamefont {Z.}~\bibnamefont
  {Lu}}\ and\ \bibinfo {author} {\bibfnamefont {O.}~\bibnamefont {Raz}},\
  }\bibfield  {title} {\bibinfo {title} {Nonequilibrium thermodynamics of the
  {Markovian Mpemba} effect and its inverse},\ }\href
  {https://doi.org/10.1073/pnas.1701264114} {\bibfield  {journal} {\bibinfo
  {journal} {Proc. Natl. Acad. Sci. U.S.A.}\ }\textbf {\bibinfo {volume}
  {114}},\ \bibinfo {pages} {5083} (\bibinfo {year} {2017})}\BibitemShut
  {NoStop}%
\bibitem [{\citenamefont {Lasanta}\ \emph {et~al.}(2017)\citenamefont
  {Lasanta}, \citenamefont {Vega~Reyes}, \citenamefont {Prados},\ and\
  \citenamefont {Santos}}]{Lasanta2017}%
  \BibitemOpen
  \bibfield  {author} {\bibinfo {author} {\bibfnamefont {A.}~\bibnamefont
  {Lasanta}}, \bibinfo {author} {\bibfnamefont {F.}~\bibnamefont {Vega~Reyes}},
  \bibinfo {author} {\bibfnamefont {A.}~\bibnamefont {Prados}},\ and\ \bibinfo
  {author} {\bibfnamefont {A.}~\bibnamefont {Santos}},\ }\bibfield  {title}
  {\bibinfo {title} {When the hotter cools more quickly: Mpemba effect in
  granular fluids},\ }\href {https://doi.org/10.1103/PhysRevLett.119.148001}
  {\bibfield  {journal} {\bibinfo  {journal} {Phys. Rev. Lett.}\ }\textbf
  {\bibinfo {volume} {119}},\ \bibinfo {pages} {148001} (\bibinfo {year}
  {2017})}\BibitemShut {NoStop}%
\bibitem [{\citenamefont {Carollo}\ \emph {et~al.}(2021)\citenamefont
  {Carollo}, \citenamefont {Lasanta},\ and\ \citenamefont
  {Lesanovsky}}]{Carollo2021}%
  \BibitemOpen
  \bibfield  {author} {\bibinfo {author} {\bibfnamefont {F.}~\bibnamefont
  {Carollo}}, \bibinfo {author} {\bibfnamefont {A.}~\bibnamefont {Lasanta}},\
  and\ \bibinfo {author} {\bibfnamefont {I.}~\bibnamefont {Lesanovsky}},\
  }\bibfield  {title} {\bibinfo {title} {Exponentially accelerated approach to
  stationarity in {Markovian} open quantum systems through the {Mpemba}
  effect},\ }\href {https://doi.org/10.1103/PhysRevLett.127.060401} {\bibfield
  {journal} {\bibinfo  {journal} {Phys. Rev. Lett.}\ }\textbf {\bibinfo
  {volume} {127}},\ \bibinfo {pages} {060401} (\bibinfo {year}
  {2021})}\BibitemShut {NoStop}%
\bibitem [{\citenamefont {Kochsiek}\ \emph {et~al.}(2022)\citenamefont
  {Kochsiek}, \citenamefont {Carollo},\ and\ \citenamefont
  {Lesanovsky}}]{Kochsiek2022}%
  \BibitemOpen
  \bibfield  {author} {\bibinfo {author} {\bibfnamefont {S.}~\bibnamefont
  {Kochsiek}}, \bibinfo {author} {\bibfnamefont {F.}~\bibnamefont {Carollo}},\
  and\ \bibinfo {author} {\bibfnamefont {I.}~\bibnamefont {Lesanovsky}},\
  }\bibfield  {title} {\bibinfo {title} {Accelerating the approach of
  dissipative quantum spin systems towards stationarity through global spin
  rotations},\ }\href {https://doi.org/10.1103/PhysRevA.106.012207} {\bibfield
  {journal} {\bibinfo  {journal} {Phys. Rev. A}\ }\textbf {\bibinfo {volume}
  {106}},\ \bibinfo {pages} {012207} (\bibinfo {year} {2022})}\BibitemShut
  {NoStop}%
\bibitem [{\citenamefont {Chatterjee}\ \emph {et~al.}(2023)\citenamefont
  {Chatterjee}, \citenamefont {Takada},\ and\ \citenamefont
  {Hayakawa}}]{Chatterjee2023}%
  \BibitemOpen
  \bibfield  {author} {\bibinfo {author} {\bibfnamefont {A.~K.}\ \bibnamefont
  {Chatterjee}}, \bibinfo {author} {\bibfnamefont {S.}~\bibnamefont {Takada}},\
  and\ \bibinfo {author} {\bibfnamefont {H.}~\bibnamefont {Hayakawa}},\
  }\bibfield  {title} {\bibinfo {title} {Quantum {Mpemba} effect in a quantum
  dot with reservoirs},\ }\href
  {https://doi.org/10.1103/PhysRevLett.131.080402} {\bibfield  {journal}
  {\bibinfo  {journal} {Phys. Rev. Lett.}\ }\textbf {\bibinfo {volume} {131}},\
  \bibinfo {pages} {080402} (\bibinfo {year} {2023})}\BibitemShut {NoStop}%
\bibitem [{\citenamefont {Moroder}\ \emph {et~al.}(2024)\citenamefont
  {Moroder}, \citenamefont {Culhane}, \citenamefont {Zawadzki},\ and\
  \citenamefont {Goold}}]{Moroder2024}%
  \BibitemOpen
  \bibfield  {author} {\bibinfo {author} {\bibfnamefont {M.}~\bibnamefont
  {Moroder}}, \bibinfo {author} {\bibfnamefont {O.}~\bibnamefont {Culhane}},
  \bibinfo {author} {\bibfnamefont {K.}~\bibnamefont {Zawadzki}},\ and\
  \bibinfo {author} {\bibfnamefont {J.}~\bibnamefont {Goold}},\ }\bibfield
  {title} {\bibinfo {title} {Thermodynamics of the quantum mpemba effect},\
  }\href {https://doi.org/10.1103/PhysRevLett.133.140404} {\bibfield  {journal}
  {\bibinfo  {journal} {Phys. Rev. Lett.}\ }\textbf {\bibinfo {volume} {133}},\
  \bibinfo {pages} {140404} (\bibinfo {year} {2024})}\BibitemShut {NoStop}%
\bibitem [{\citenamefont {Joshi}\ \emph {et~al.}(2024)\citenamefont {Joshi},
  \citenamefont {Franke}, \citenamefont {Rath}, \citenamefont {Ares},
  \citenamefont {Murciano}, \citenamefont {Kranzl}, \citenamefont {Blatt},
  \citenamefont {Zoller}, \citenamefont {Vermersch}, \citenamefont {Calabrese},
  \citenamefont {Roos},\ and\ \citenamefont {Joshi}}]{Joshi2024}%
  \BibitemOpen
  \bibfield  {author} {\bibinfo {author} {\bibfnamefont {L.~K.}\ \bibnamefont
  {Joshi}}, \bibinfo {author} {\bibfnamefont {J.}~\bibnamefont {Franke}},
  \bibinfo {author} {\bibfnamefont {A.}~\bibnamefont {Rath}}, \bibinfo {author}
  {\bibfnamefont {F.}~\bibnamefont {Ares}}, \bibinfo {author} {\bibfnamefont
  {S.}~\bibnamefont {Murciano}}, \bibinfo {author} {\bibfnamefont
  {F.}~\bibnamefont {Kranzl}}, \bibinfo {author} {\bibfnamefont
  {R.}~\bibnamefont {Blatt}}, \bibinfo {author} {\bibfnamefont
  {P.}~\bibnamefont {Zoller}}, \bibinfo {author} {\bibfnamefont
  {B.}~\bibnamefont {Vermersch}}, \bibinfo {author} {\bibfnamefont
  {P.}~\bibnamefont {Calabrese}}, \bibinfo {author} {\bibfnamefont {C.~F.}\
  \bibnamefont {Roos}},\ and\ \bibinfo {author} {\bibfnamefont {M.~K.}\
  \bibnamefont {Joshi}},\ }\bibfield  {title} {\bibinfo {title} {Observing the
  quantum mpemba effect in quantum simulations},\ }\href
  {https://doi.org/10.1103/PhysRevLett.133.010402} {\bibfield  {journal}
  {\bibinfo  {journal} {Phys. Rev. Lett.}\ }\textbf {\bibinfo {volume} {133}},\
  \bibinfo {pages} {010402} (\bibinfo {year} {2024})}\BibitemShut {NoStop}%
\bibitem [{\citenamefont {Aharony~Shapira}\ \emph {et~al.}(2024)\citenamefont
  {Aharony~Shapira}, \citenamefont {Shapira}, \citenamefont {Markov},
  \citenamefont {Teza}, \citenamefont {Akerman}, \citenamefont {Raz},\ and\
  \citenamefont {Ozeri}}]{Shapira2024}%
  \BibitemOpen
  \bibfield  {author} {\bibinfo {author} {\bibfnamefont {S.}~\bibnamefont
  {Aharony~Shapira}}, \bibinfo {author} {\bibfnamefont {Y.}~\bibnamefont
  {Shapira}}, \bibinfo {author} {\bibfnamefont {J.}~\bibnamefont {Markov}},
  \bibinfo {author} {\bibfnamefont {G.}~\bibnamefont {Teza}}, \bibinfo {author}
  {\bibfnamefont {N.}~\bibnamefont {Akerman}}, \bibinfo {author} {\bibfnamefont
  {O.}~\bibnamefont {Raz}},\ and\ \bibinfo {author} {\bibfnamefont
  {R.}~\bibnamefont {Ozeri}},\ }\bibfield  {title} {\bibinfo {title} {Inverse
  mpemba effect demonstrated on a single trapped ion qubit},\ }\href
  {https://doi.org/10.1103/PhysRevLett.133.010403} {\bibfield  {journal}
  {\bibinfo  {journal} {Phys. Rev. Lett.}\ }\textbf {\bibinfo {volume} {133}},\
  \bibinfo {pages} {010403} (\bibinfo {year} {2024})}\BibitemShut {NoStop}%
\bibitem [{\citenamefont {Zhang}\ \emph {et~al.}(2025)\citenamefont {Zhang},
  \citenamefont {Xia}, \citenamefont {Wu}, \citenamefont {Chen}, \citenamefont
  {Zhang}, \citenamefont {Xie}, \citenamefont {Su}, \citenamefont {Wu},
  \citenamefont {Qiu}, \citenamefont {Chen}, \citenamefont {Li}, \citenamefont
  {Jing},\ and\ \citenamefont {Zhou}}]{Zhang2025}%
  \BibitemOpen
  \bibfield  {author} {\bibinfo {author} {\bibfnamefont {J.}~\bibnamefont
  {Zhang}}, \bibinfo {author} {\bibfnamefont {G.}~\bibnamefont {Xia}}, \bibinfo
  {author} {\bibfnamefont {C.-W.}\ \bibnamefont {Wu}}, \bibinfo {author}
  {\bibfnamefont {T.}~\bibnamefont {Chen}}, \bibinfo {author} {\bibfnamefont
  {Q.}~\bibnamefont {Zhang}}, \bibinfo {author} {\bibfnamefont
  {Y.}~\bibnamefont {Xie}}, \bibinfo {author} {\bibfnamefont {W.-B.}\
  \bibnamefont {Su}}, \bibinfo {author} {\bibfnamefont {W.}~\bibnamefont {Wu}},
  \bibinfo {author} {\bibfnamefont {C.-W.}\ \bibnamefont {Qiu}}, \bibinfo
  {author} {\bibfnamefont {P.-X.}\ \bibnamefont {Chen}}, \bibinfo {author}
  {\bibfnamefont {W.}~\bibnamefont {Li}}, \bibinfo {author} {\bibfnamefont
  {H.}~\bibnamefont {Jing}},\ and\ \bibinfo {author} {\bibfnamefont {Y.-L.}\
  \bibnamefont {Zhou}},\ }\bibfield  {title} {\bibinfo {title} {Observation of
  quantum strong mpemba effect},\ }\href
  {https://doi.org/10.1038/s41467-024-54303-0} {\bibfield  {journal} {\bibinfo
  {journal} {Nature Commun.}\ }\textbf {\bibinfo {volume} {16}},\ \bibinfo
  {pages} {301} (\bibinfo {year} {2025})}\BibitemShut {NoStop}%
\bibitem [{\citenamefont {Chatterjee}\ \emph {et~al.}(2025)\citenamefont
  {Chatterjee}, \citenamefont {Khan}, \citenamefont {Jain},\ and\ \citenamefont
  {Mahesh}}]{Chatterjee2025}%
  \BibitemOpen
  \bibfield  {author} {\bibinfo {author} {\bibfnamefont {A.}~\bibnamefont
  {Chatterjee}}, \bibinfo {author} {\bibfnamefont {S.}~\bibnamefont {Khan}},
  \bibinfo {author} {\bibfnamefont {S.}~\bibnamefont {Jain}},\ and\ \bibinfo
  {author} {\bibfnamefont {T.~S.}\ \bibnamefont {Mahesh}},\ }\bibfield  {title}
  {\bibinfo {title} {Direct experimental observation of quantum mpemba effect
  without bath engineeringt},\ }\href@noop {} {\bibfield  {journal} {\bibinfo
  {journal} {arXiv:2509.13451}\ } (\bibinfo {year} {2025})}\BibitemShut
  {NoStop}%
\bibitem [{\citenamefont {Schnepper}\ \emph {et~al.}(2025)\citenamefont
  {Schnepper}, \citenamefont {de~Oliveira}, \citenamefont {Vieira},
  \citenamefont {Zawadzki},\ and\ \citenamefont {Serra}}]{Schnepper2025}%
  \BibitemOpen
  \bibfield  {author} {\bibinfo {author} {\bibfnamefont {B.~P.}\ \bibnamefont
  {Schnepper}}, \bibinfo {author} {\bibfnamefont {J.~L.~D.}\ \bibnamefont
  {de~Oliveira}}, \bibinfo {author} {\bibfnamefont {C.~H.~S.}\ \bibnamefont
  {Vieira}}, \bibinfo {author} {\bibfnamefont {K.}~\bibnamefont {Zawadzki}},\
  and\ \bibinfo {author} {\bibfnamefont {R.~M.}\ \bibnamefont {Serra}},\
  }\bibfield  {title} {\bibinfo {title} {Experimental observation and
  application of the genuine quantum mpemba effect},\ }\href@noop {} {\bibfield
   {journal} {\bibinfo  {journal} {arXiv:2511.14552}\ } (\bibinfo {year}
  {2025})}\BibitemShut {NoStop}%
\bibitem [{\citenamefont {Strachan}\ \emph {et~al.}(2025)\citenamefont
  {Strachan}, \citenamefont {Purkayastha},\ and\ \citenamefont
  {Clark}}]{Strachan2025}%
  \BibitemOpen
  \bibfield  {author} {\bibinfo {author} {\bibfnamefont {D.~J.}\ \bibnamefont
  {Strachan}}, \bibinfo {author} {\bibfnamefont {A.}~\bibnamefont
  {Purkayastha}},\ and\ \bibinfo {author} {\bibfnamefont {S.~R.}\ \bibnamefont
  {Clark}},\ }\bibfield  {title} {\bibinfo {title} {Non-markovian quantum
  mpemba effect},\ }\href {https://doi.org/10.1103/PhysRevLett.134.220403}
  {\bibfield  {journal} {\bibinfo  {journal} {Phys. Rev. Lett.}\ }\textbf
  {\bibinfo {volume} {134}},\ \bibinfo {pages} {220403} (\bibinfo {year}
  {2025})}\BibitemShut {NoStop}%
\bibitem [{\citenamefont {Vu}\ and\ \citenamefont {Hayakawa}(2025)}]{Vu2025}%
  \BibitemOpen
  \bibfield  {author} {\bibinfo {author} {\bibfnamefont {T.~V.}\ \bibnamefont
  {Vu}}\ and\ \bibinfo {author} {\bibfnamefont {H.}~\bibnamefont {Hayakawa}},\
  }\bibfield  {title} {\bibinfo {title} {Thermomajorization {Mpemba} effect},\
  }\href {https://doi.org/10.1103/PhysRevLett.134.107101} {\bibfield  {journal}
  {\bibinfo  {journal} {Phys. Rev. Lett.}\ }\textbf {\bibinfo {volume} {134}},\
  \bibinfo {pages} {107101} (\bibinfo {year} {2025})}\BibitemShut {NoStop}%
\bibitem [{\citenamefont {Choquehuanca}\ \emph {et~al.}(2025)\citenamefont
  {Choquehuanca}, \citenamefont {Obando}, \citenamefont {Sarandy},\ and\
  \citenamefont {de~Paula}}]{Choquehuanca2025}%
  \BibitemOpen
  \bibfield  {author} {\bibinfo {author} {\bibfnamefont {J.~M.~Z.}\
  \bibnamefont {Choquehuanca}}, \bibinfo {author} {\bibfnamefont {P.~A.~C.}\
  \bibnamefont {Obando}}, \bibinfo {author} {\bibfnamefont {M.~S.}\
  \bibnamefont {Sarandy}},\ and\ \bibinfo {author} {\bibfnamefont {F.~M.}\
  \bibnamefont {de~Paula}},\ }\bibfield  {title} {\bibinfo {title}
  {Ergotropy-based quantum thermodynamics},\ }\href
  {https://doi.org/10.1103/3vt1-m8z2} {\bibfield  {journal} {\bibinfo
  {journal} {Phys. Rev. A}\ }\textbf {\bibinfo {volume} {112}},\ \bibinfo
  {pages} {052220} (\bibinfo {year} {2025})}\BibitemShut {NoStop}%
\bibitem [{\citenamefont {Lindblad}(1976)}]{Lindblad1976}%
  \BibitemOpen
  \bibfield  {author} {\bibinfo {author} {\bibfnamefont {G.}~\bibnamefont
  {Lindblad}},\ }\bibfield  {title} {\bibinfo {title} {On the generators of
  quantum dynamical semigroups},\ }\href {https://doi.org/10.1007/BF01608499}
  {\bibfield  {journal} {\bibinfo  {journal} {Commun. Math. Phys.}\ }\textbf
  {\bibinfo {volume} {48}},\ \bibinfo {pages} {119} (\bibinfo {year}
  {1976})}\BibitemShut {NoStop}%
\bibitem [{\citenamefont {Gorini}\ \emph {et~al.}(1976)\citenamefont {Gorini},
  \citenamefont {Kossakowski},\ and\ \citenamefont {Sudarshan}}]{Gorini1976}%
  \BibitemOpen
  \bibfield  {author} {\bibinfo {author} {\bibfnamefont {V.}~\bibnamefont
  {Gorini}}, \bibinfo {author} {\bibfnamefont {A.}~\bibnamefont
  {Kossakowski}},\ and\ \bibinfo {author} {\bibfnamefont {E.~C.~G.}\
  \bibnamefont {Sudarshan}},\ }\bibfield  {title} {\bibinfo {title} {Completely
  positive dynamical semigroups of {N}-level systems},\ }\href
  {https://doi.org/10.1063/1.522979} {\bibfield  {journal} {\bibinfo  {journal}
  {J. Math. Phys.}\ }\textbf {\bibinfo {volume} {17}},\ \bibinfo {pages} {821}
  (\bibinfo {year} {1976})}\BibitemShut {NoStop}%
\bibitem [{\citenamefont {Breuer}\ and\ \citenamefont
  {Petruccione}(2002)}]{Breuer2002}%
  \BibitemOpen
  \bibfield  {author} {\bibinfo {author} {\bibfnamefont {H.-P.}\ \bibnamefont
  {Breuer}}\ and\ \bibinfo {author} {\bibfnamefont {F.}~\bibnamefont
  {Petruccione}},\ }\href@noop {} {\emph {\bibinfo {title} {The Theory of Open
  Quantum Systems}}}\ (\bibinfo  {publisher} {Oxford University Press},\
  \bibinfo {year} {2002})\BibitemShut {NoStop}%
\bibitem [{\citenamefont {Sarandy}\ and\ \citenamefont
  {Lidar}(2005)}]{Sarandy2005}%
  \BibitemOpen
  \bibfield  {author} {\bibinfo {author} {\bibfnamefont {M.~S.}\ \bibnamefont
  {Sarandy}}\ and\ \bibinfo {author} {\bibfnamefont {D.~A.}\ \bibnamefont
  {Lidar}},\ }\bibfield  {title} {\bibinfo {title} {Adiabatic approximation in
  open quantum systems},\ }\href {https://doi.org/10.1103/PhysRevA.71.012331}
  {\bibfield  {journal} {\bibinfo  {journal} {Phys. Rev. A}\ }\textbf {\bibinfo
  {volume} {71}},\ \bibinfo {pages} {012331} (\bibinfo {year}
  {2005})}\BibitemShut {NoStop}%
\bibitem [{\citenamefont {Santos}\ and\ \citenamefont
  {Sarandy}(2020)}]{Santos2020}%
  \BibitemOpen
  \bibfield  {author} {\bibinfo {author} {\bibfnamefont {A.~C.}\ \bibnamefont
  {Santos}}\ and\ \bibinfo {author} {\bibfnamefont {M.~S.}\ \bibnamefont
  {Sarandy}},\ }\bibfield  {title} {\bibinfo {title} {Sufficient conditions for
  adiabaticity in open quantum systems},\ }\href
  {https://doi.org/10.1103/PhysRevA.102.052215} {\bibfield  {journal} {\bibinfo
   {journal} {Phys. Rev. A}\ }\textbf {\bibinfo {volume} {102}},\ \bibinfo
  {pages} {052215} (\bibinfo {year} {2020})}\BibitemShut {NoStop}%
\bibitem [{\citenamefont {Davies}(1974)}]{Davies1974}%
  \BibitemOpen
  \bibfield  {author} {\bibinfo {author} {\bibfnamefont {E.~B.}\ \bibnamefont
  {Davies}},\ }\bibfield  {title} {\bibinfo {title} {Markovian master
  equations},\ }\href@noop {} {\bibfield  {journal} {\bibinfo  {journal}
  {Commun. Math. Phys.}\ }\textbf {\bibinfo {volume} {39}},\ \bibinfo {pages}
  {91} (\bibinfo {year} {1974})}\BibitemShut {NoStop}%
\bibitem [{\citenamefont {Sapui}\ \emph {et~al.}(2026)\citenamefont {Sapui},
  \citenamefont {Konar},\ and\ \citenamefont {Sen~De}}]{Sapui2026}%
  \BibitemOpen
  \bibfield  {author} {\bibinfo {author} {\bibfnamefont {T.}~\bibnamefont
  {Sapui}}, \bibinfo {author} {\bibfnamefont {T.~K.}\ \bibnamefont {Konar}},\
  and\ \bibinfo {author} {\bibfnamefont {A.}~\bibnamefont {Sen~De}},\
  }\bibfield  {title} {\bibinfo {title} {Ergotropic mpemba crossings in
  finite-dimensional quantum batteries},\ }\href@noop {} {\bibfield  {journal}
  {\bibinfo  {journal} {arXiv:2602.11056v2}\ } (\bibinfo {year}
  {2026})}\BibitemShut {NoStop}%
\bibitem [{\citenamefont {Nielsen}\ and\ \citenamefont
  {Chuang}(2000)}]{nielsen2000}%
  \BibitemOpen
  \bibfield  {author} {\bibinfo {author} {\bibfnamefont {M.~A.}\ \bibnamefont
  {Nielsen}}\ and\ \bibinfo {author} {\bibfnamefont {I.~L.}\ \bibnamefont
  {Chuang}},\ }\href@noop {} {\emph {\bibinfo {title} {Quantum Computation and
  Quantum Information}}}\ (\bibinfo  {publisher} {Cambridge University Press},\
  \bibinfo {year} {2000})\BibitemShut {NoStop}%
\bibitem [{\citenamefont {Baumgratz}\ \emph {et~al.}(2014)\citenamefont
  {Baumgratz}, \citenamefont {Cramer},\ and\ \citenamefont
  {Plenio}}]{Baumgratz2014}%
  \BibitemOpen
  \bibfield  {author} {\bibinfo {author} {\bibfnamefont {T.}~\bibnamefont
  {Baumgratz}}, \bibinfo {author} {\bibfnamefont {M.}~\bibnamefont {Cramer}},\
  and\ \bibinfo {author} {\bibfnamefont {M.~B.}\ \bibnamefont {Plenio}},\
  }\bibfield  {title} {\bibinfo {title} {Quantifying coherence},\ }\href
  {https://doi.org/10.1103/PhysRevLett.113.140401} {\bibfield  {journal}
  {\bibinfo  {journal} {Phys. Rev. Lett.}\ }\textbf {\bibinfo {volume} {113}},\
  \bibinfo {pages} {140401} (\bibinfo {year} {2014})}\BibitemShut {NoStop}%
\bibitem [{\citenamefont {Francica}\ \emph {et~al.}(2020)\citenamefont
  {Francica}, \citenamefont {Binder}, \citenamefont {Guarnieri}, \citenamefont
  {Mitchison}, \citenamefont {Goold},\ and\ \citenamefont
  {Plastina}}]{Francica2020}%
  \BibitemOpen
  \bibfield  {author} {\bibinfo {author} {\bibfnamefont {G.}~\bibnamefont
  {Francica}}, \bibinfo {author} {\bibfnamefont {F.~C.}\ \bibnamefont
  {Binder}}, \bibinfo {author} {\bibfnamefont {G.}~\bibnamefont {Guarnieri}},
  \bibinfo {author} {\bibfnamefont {M.~T.}\ \bibnamefont {Mitchison}}, \bibinfo
  {author} {\bibfnamefont {J.}~\bibnamefont {Goold}},\ and\ \bibinfo {author}
  {\bibfnamefont {F.}~\bibnamefont {Plastina}},\ }\bibfield  {title} {\bibinfo
  {title} {Quantum coherence and ergotropy},\ }\href
  {https://doi.org/10.1103/PhysRevLett.125.180603} {\bibfield  {journal}
  {\bibinfo  {journal} {Phys. Rev. Lett.}\ }\textbf {\bibinfo {volume} {125}},\
  \bibinfo {pages} {180603} (\bibinfo {year} {2020})}\BibitemShut {NoStop}%
\bibitem [{\citenamefont {Choquehuanca}\ \emph {et~al.}(2024)\citenamefont
  {Choquehuanca}, \citenamefont {Obando}, \citenamefont {de~Paula},\ and\
  \citenamefont {Sarandy}}]{Choquehuanca2024}%
  \BibitemOpen
  \bibfield  {author} {\bibinfo {author} {\bibfnamefont {J.~M.~Z.}\
  \bibnamefont {Choquehuanca}}, \bibinfo {author} {\bibfnamefont {P.~A.~C.}\
  \bibnamefont {Obando}}, \bibinfo {author} {\bibfnamefont {F.~M.}\
  \bibnamefont {de~Paula}},\ and\ \bibinfo {author} {\bibfnamefont {M.~S.}\
  \bibnamefont {Sarandy}},\ }\bibfield  {title} {\bibinfo {title} {Qubit
  dynamics of ergotropy and environment-induced work},\ }\href
  {https://doi.org/10.1103/PhysRevA.109.052219} {\bibfield  {journal} {\bibinfo
   {journal} {Phys. Rev. A}\ }\textbf {\bibinfo {volume} {109}},\ \bibinfo
  {pages} {052219} (\bibinfo {year} {2024})}\BibitemShut {NoStop}%
\end{thebibliography}

%

\end{document}